\documentclass[prl,aps,twocolumn,singlespace]{revtex4-1}
\usepackage{bm}
\usepackage{graphicx}
\usepackage{amsfonts}
\usepackage{amsmath}
\usepackage{amssymb}
\newcommand {\be}{\begin{equation}}
\newcommand {\ee}{\end{equation}}
\newcommand {\bea}{\begin{eqnarray}}
\newcommand {\eea}{\end{eqnarray}}

\begin{document}
\linespread{1.0}
\title{Fractal nanostructures with the Hilbert curve geometry as a SERS substrate}
\author{Ilya Grigorenko}
\affiliation{Penn State}
\date{\today}

\begin{abstract}
We suggest a new type of substrates for the Surface Enhanced Raman Scattering measurements with the geometry
based on  self-similar fractal space filling curves. 
As an example, we have studied theoretically the dielectric response properties of  
doped semiconductor nanostructures, where the conducting electrons are trapped in the effective 
potential having the geometry of the Hilbert curve. We have found that the system 
may exhibit the induced charge distribution 
specific for either  two dimensional or one dimensional systems, depending on the frequency of the external applied field. 
We have demonstrated that with the increasing of the depth of the trapping potential the resonance of the system 
counterintuitively
shifts to lower frequencies. 
\end{abstract}

\maketitle
Quick, reliable and selective single molecule detection of different molecular species is a challenging problem.
The Surface Enhanced Raman Scattering (SERS) technique is considered as one of the most powerful methods for single molecular detection  \cite{shalaev,sers}.
The local electromagnetic field enhancement on random metallic substrates of irregular, fractal-like geometries, or on corrugated metal surfaces 
was considered as one of the primary enhancement 
mechanisms for the SERS measurements since the discovery of the phenomenon. 
Such geometry results in a local field intensity enhancement by a factor up to $|E|^2\propto10^3-10^5$ in so called ''hot spots`` \cite{stockman,stockman1}. 
The Raman signal is proportional to $|E|^4$, therefore a molecule in a ''hot spot`` may emit up to $10^6-10^{10}$ times stronger signal.
Such kind of substrates were intensively studied both experimentally  and  theoretically \cite{sers,shalaev,raman_frac1,raman_frac2,raman_frac3}.
For example, it was shown, that the statistics of ``hot spots'' in such substrates may be governed by power law distributions \cite{stockman}. 
However,  random substrates have significant disadvantage of poor reproducibility of measurements for different substrate samples.
Another disadvantage of these type  of substrate geometries is a relatively low degree of controllability of the resonance properties of the substrate.
     
In the last decade there have been a considerable interest in stable, reproducible SERS substrates. For example, arrays of 
nanoparticle pairs \cite{nurmikko,ilya2} 
or  photonic/plasmonic crystal structures were considered recently as an alternative geometry with more degrees of 
control of the substrate's resonance properties \cite{lithography_plasmonic_sensors}. Such kind of designs can generate a periodic
 pattern of a high intensity 
``hot spots'', however, 
a relatively large fraction of the area is SERS inactive.
In this paper we consider a new type of geometry for SERS substrates, which combines the advantages  of irregular 
fractal substrates, resulting in a strong local field enhancement, and  periodic plasmonic structures, characterized by relatively easy 
reproducibility and a high degree of controllability of the substrate's resonance properties.
We suggest to use for the SERS substrate a geometry derived from a special class of fractal-like space filling curves \cite{curves}. 
For example, the Hilbert curve  is a special case of space filling curves.
 It has a large number of “turns” and “corners” for a given length of a curve (please, see Figs.(1,2)). 
Note, that the Hilbert curve is a self similar structure, which 
may be built making a finite (or infinite) number of iterations. 
A nanoscale SERS substrate will not be a true fractal because of the lack of an 
infinitely fine spatial resolution. In particular, at the atomic scale the quantization of electronic 
states will start to play a significant role.
 Since the local field enhancement is the most significant at surfaces with 
a large local curvature (at the corners), the Hilbert curve  is a good candidate as a substrate for the SERS detection
with a high probability for a molecule to
 hit a ''hot spot”. Another advantage of the Hilbert curve is that while it has a self-similar fractal structure, 
its Hausdorff dimension is 2,
 i.e. the Hilbert curve uniformly covers surface without empty spots.
Regular fractal-like structures have been considered for nano-photonic applications before \cite{serp,spie,antenna}.
However, in these works the authors considered fractal structures with the Hausdorff dimension  $1<D<2$, and therefore, such structures
do not fill  surface uniformly. 
 
Space filling  substrates can be manufactured using modern lithography techniques \cite{lithography_plasmonic_sensors}.  A possible approach is to use ion beam lithography methods to etch trenches with the shape
of the Hilbert curve on a metal surface or a thin film. In a case of a thin film one may vary the depth of the 
trenches, continuously moving from two dimensional geometry of a  thin film to the 
one dimensional self-similar geometry of a Hilbert wire.
Another alternative approach is to use metal nanospheres organized on a surface as the Hilbert curve. 
In the last case 
one has an additional degrees of control of the resonance properties.
 By choosing certain spheres radii and the inter-sphere distance one may shift the resonances 
in the most suitable range 
for the single molecule detection of a given chemical specie. In order to 
increase the degree of control of the resonant properties of the system one may fabricate 
the nanostructure using spheres made of 
 different materials (for example, silver and gold), or using bimetal spheres \cite{bimetal}. 
One may also use Surface Tunneling Microscope techniques \cite{manoharan}
to assemble and measure the response of an atomic scale structure having the geometry of the Hilbert curve.  
In order to utilize rich dielectric properties of fractal-like 
space filling curves in modern nanoelectronics one may use photo lithography methods, 
similar to those for microchip fabrication, which will result in non-uniform doping levels in a semiconductor substrate. 
Conducting electrons will be localized in a highly non-uniform trapping potential with the Hilbert curve geometry.
Note that while the electrons will be trapped in a quasi-1D potential, the interaction with an external field will result
in the induced charge distribution having also quasi one dimensional properties. However, the  
electrons may also interact with each other via the Coulomb interaction, which will extend out of the plane of the substrate, making 
the system effectively  two dimensional. 
As we will show in this paper, the collective excitations in such systems may exhibit the induced charge distributions,
which is characteristic for either one or two dimensional systems, depending on the excitation frequency.


In order to calculate the dielectric properties of fractal nanostructures we used 
linear response theory under Random Phase Approximation (RPA)  \cite{ilya1}. This approach allows us to take into account the non-local properties of the 
dielectric response of the highly inhomogeneous atomic scale system.

 The Schr\"odinger equation for non-interacting
electrons with the effective mass $m^*_{\text e}$  moving in a
trapping potential $V(\mathbf{r})$, is given by
\begin{equation}  \label{hamiltonian}
H\Psi_i(\mathbf{r}) = \left( -\frac{\hbar^2}{2m^*_e} \nabla^2 + V(\mathbf{r})
\right) \Psi_i(\mathbf{r}) = E_i \Psi_i(\mathbf{r}).
\end{equation}
The eigenenergies $E_i$ and eigenfunctions $\Psi_i(r)$ are obtained using numerical diagonalization. 
 The
induced potential $\phi _{\text{ind}}$ is then determined from the
self-consistent integral equation \bea \label{integral_equation}
\phi _{{\text {ind}}} ({\bf r},\omega) = \int{ {\chi_0 ({\bf
r'},{\bf r''},\omega ){\phi _{{\text {tot}}} ({\bf r''},\omega)
}V_C(\left| {{\bf r} - {\bf r'}} \right|)d{ \bf r'}} d {\bf r''}}.
\eea Here $V_C(\left| {{\bf r} - {\bf r'}} \right|)$ is the Coulomb potential and 
\bea \chi_0({\bf r'},{\bf r''},\omega )=\nonumber\\
\sum_{i,j}\frac{f(E_i)-f(E_j)}{E_i-E_j-\hbar\omega-i\gamma}\Psi^*_i({\bf r'})\Psi_i({\bf r''})\Psi_j^*({\bf r''})\Psi_j({\bf r'})\eea 
is the
non-local density-density response function, $f(E_i)$ is the Fermi distribution function and $\gamma$ is the level broadening constant. Here $\phi _{\text {tot}} (%
\mathbf{r},\omega)=\phi _{\text {ext}} (\mathbf{r},\omega) +
\phi _{\text {ind}} (\textbf{r},\omega)$ is the self-consistent total
potential.  The external field is assumed to be harmonic
with frequency $\omega$, linearly polarized, and with the
wavelength much larger than the characteristic system's size, therefore $E_{ext}$ does not depend on ${\bf r}$,
 $\phi _{{\text {ind}}}({\bf r})= -|\mathbf{E}_{{\text {ext}}}| x$.
 The integral equation Eq.(2) was discretized on a real-space cubic mesh with the
lattice constant $\Delta$, and  the resulting system of
linear equations  was solved numerically.  In our simulations we used the discretization length $\Delta=1.8$nm and
we set the effective electron mass to $m^*_{\text e}=0.067m_{\text e}$  that corresponds to doped to the concentration $10^{18}$cm$^{-3}$ GaAs \cite{levi}. 
The natural energy scale $E_0$ is defined by $E_0=\hbar^2/(2 m^*_{\text e} \Delta^2)$ and  $E_0\approx174$ meV in this paper.
The  level broadening constant is set to $\gamma=1\times
10^{-3}$ $E_0$. In our simulations we assumed zero temperature $T=0$K. The dielectric response is quantified 
using the total energy of the local field  inside and outside of the nanostructure $W=\int_V|E_{\text {tot}}({\bf r})|^2d{\bf r}$, normalized by
the energy of the applied field $W_0=\int_V|E_{\text {ext}}|^2d{\bf r}$.

\begin{figure} 
\includegraphics[width=5.5cm,angle=0]{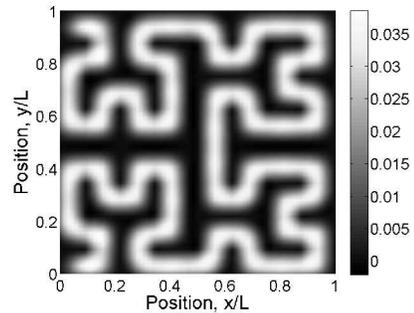}
\caption{{\protect\footnotesize {The ground state electron density for a nanostructure with the trapping potential having 
the geometry of the Hilbert curve. The size of the system $L=60$ nm. The number of electrons $N=50$. 
 }}}
\label{fig1}
\end{figure}
\begin{figure} 
\includegraphics[width=5.5cm,angle=0]{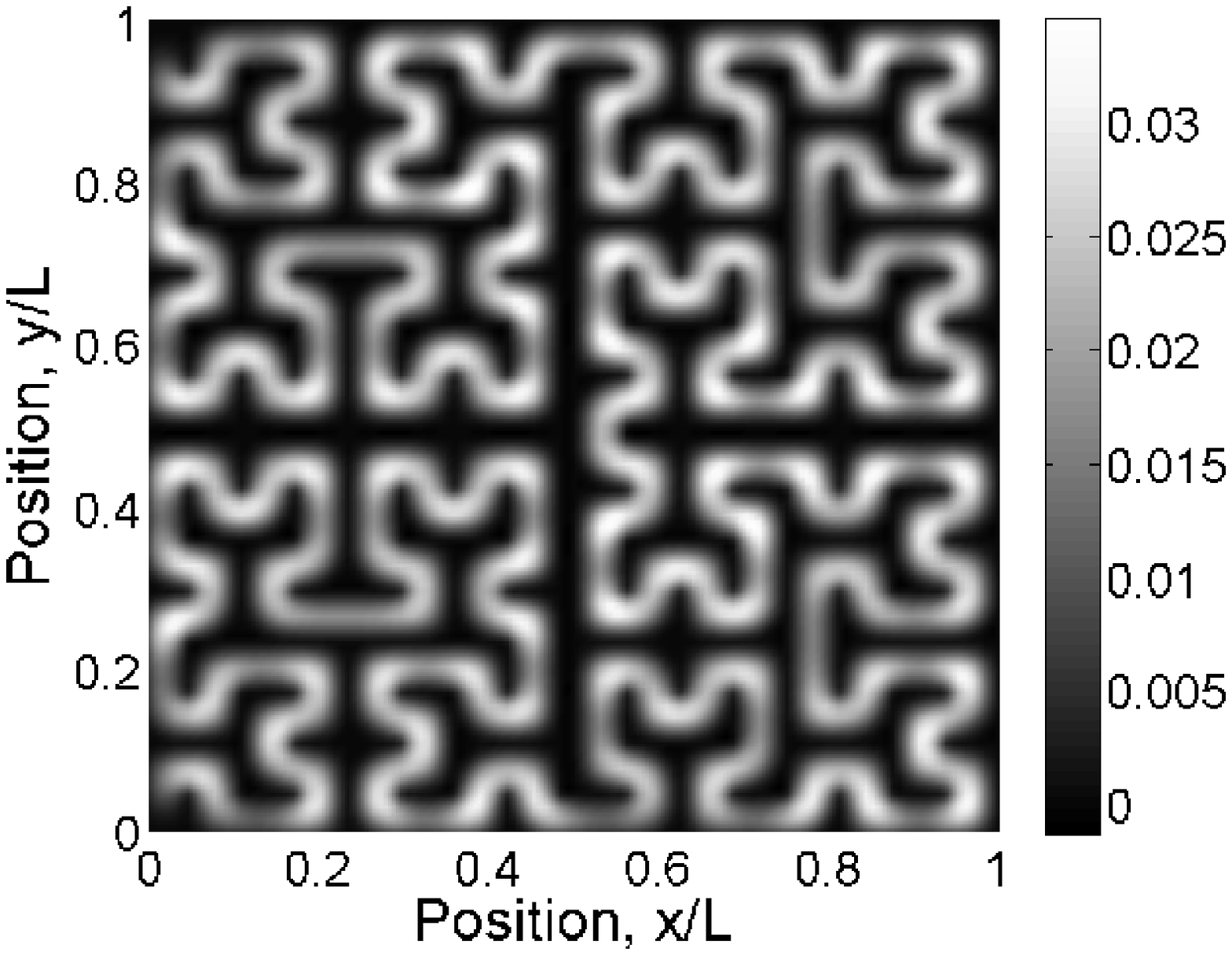}
\caption{{\protect\footnotesize {The ground state electron density for a nanostructure with the trapping potential having 
the geometry of the Hilbert curve. The size of the system $L=240$ nm.  The number of electrons $N=200$. 
 }}}
\label{fig2}
\end{figure}

\begin{figure} 

\includegraphics[width=5.5cm,angle=0]{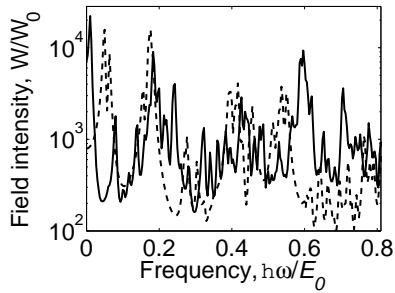}
\caption{{\protect\footnotesize { The normalised local field energy $W/W_0$ near the nanostructure shown in Fig.1 as
 a function of the frequency of the applied field $\omega$
for different number of electrons in the system: $N=25$ (solid line), $N=50$ (dashed line).}}}
\label{fig3}
\end{figure}
\begin{figure}
\includegraphics[width=5.5cm,angle=0]{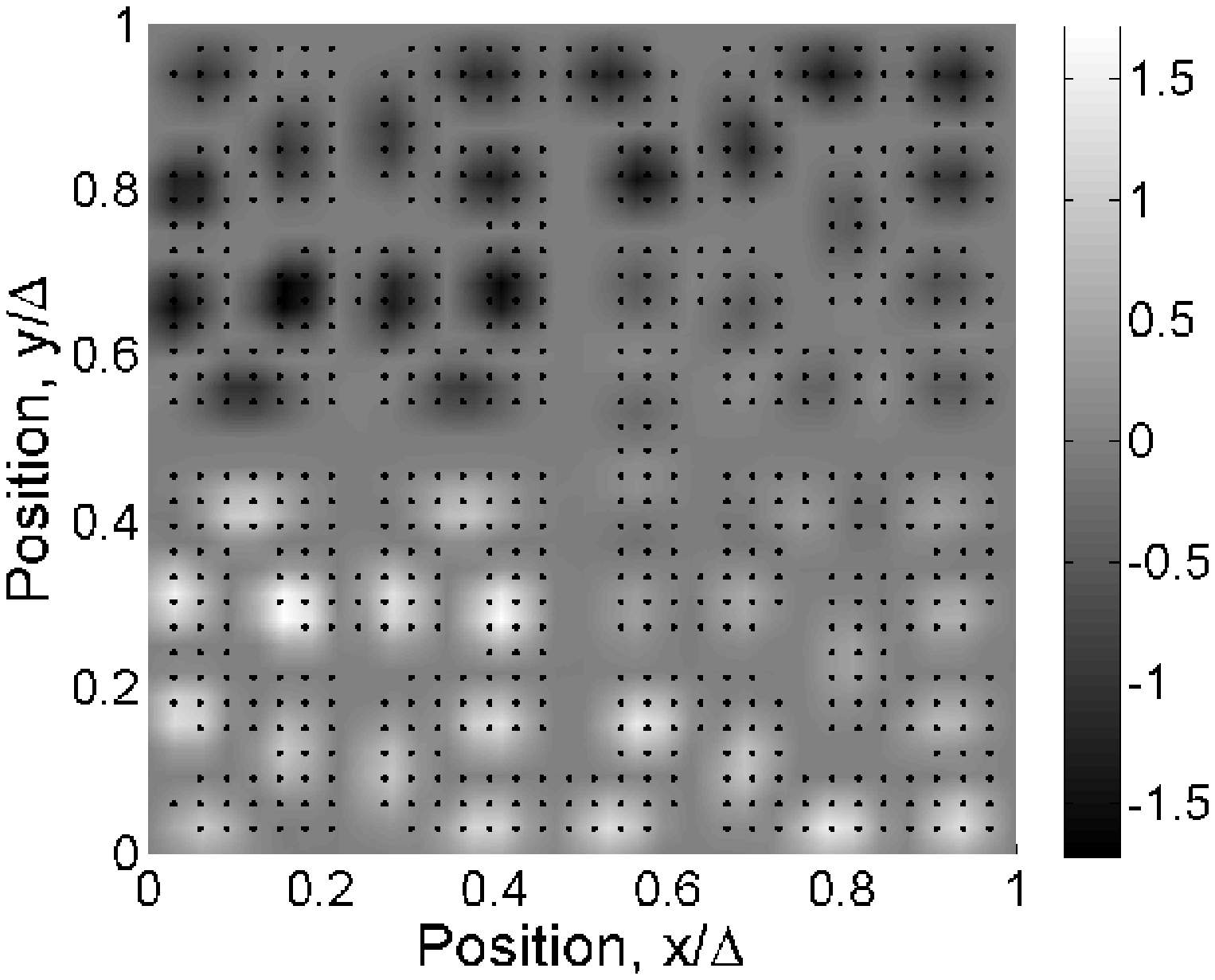}
\caption{{\protect\footnotesize {The  
induced charge density (in arb. units)  at the resonant frequency $\omega_1$.
 The direction of the  external
field $\mathbf{E}_{\text{ext}}(\protect\omega)$ points along the $y$
axis. The size of the system $L=60$ nm.  The dots mark the shape of the trapping potential.}}}
\label{fig4}
\end{figure}
\begin{figure}
\includegraphics[width=6.cm,angle=0]{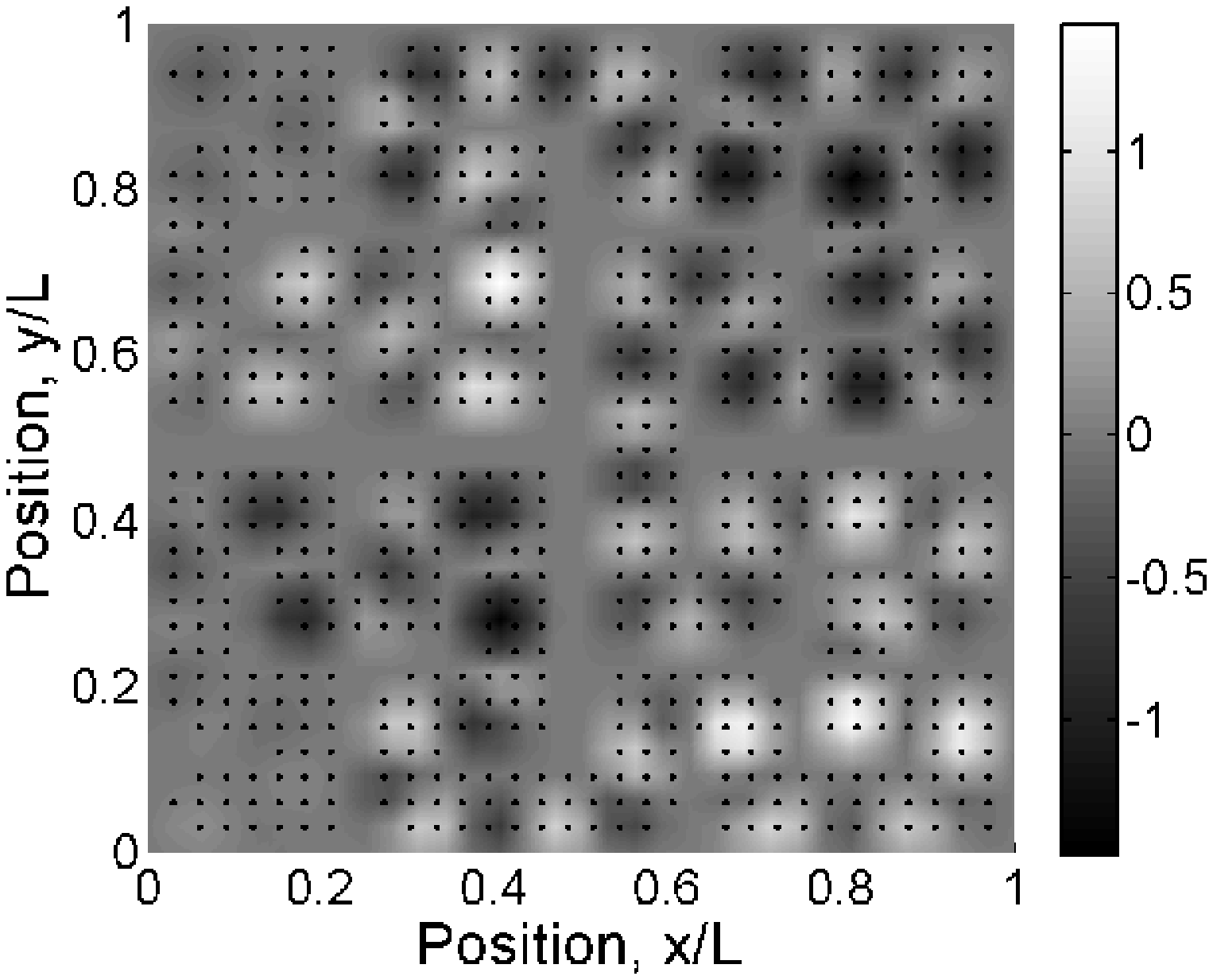}
\caption{{\protect\footnotesize {The 
induced charge density (in arb. units)  at the resonant frequency $\omega_2$.
 The direction of the  external
field $\mathbf{E}_{\text{ext}}(\protect\omega)$ points along the $y$
axis. The size of the system $L=60$ nm.  The dots mark the shape of the trapping potential.}}}
\label{fig5}
\end{figure}

\begin{figure}
\includegraphics[width=5.5cm,angle=0]{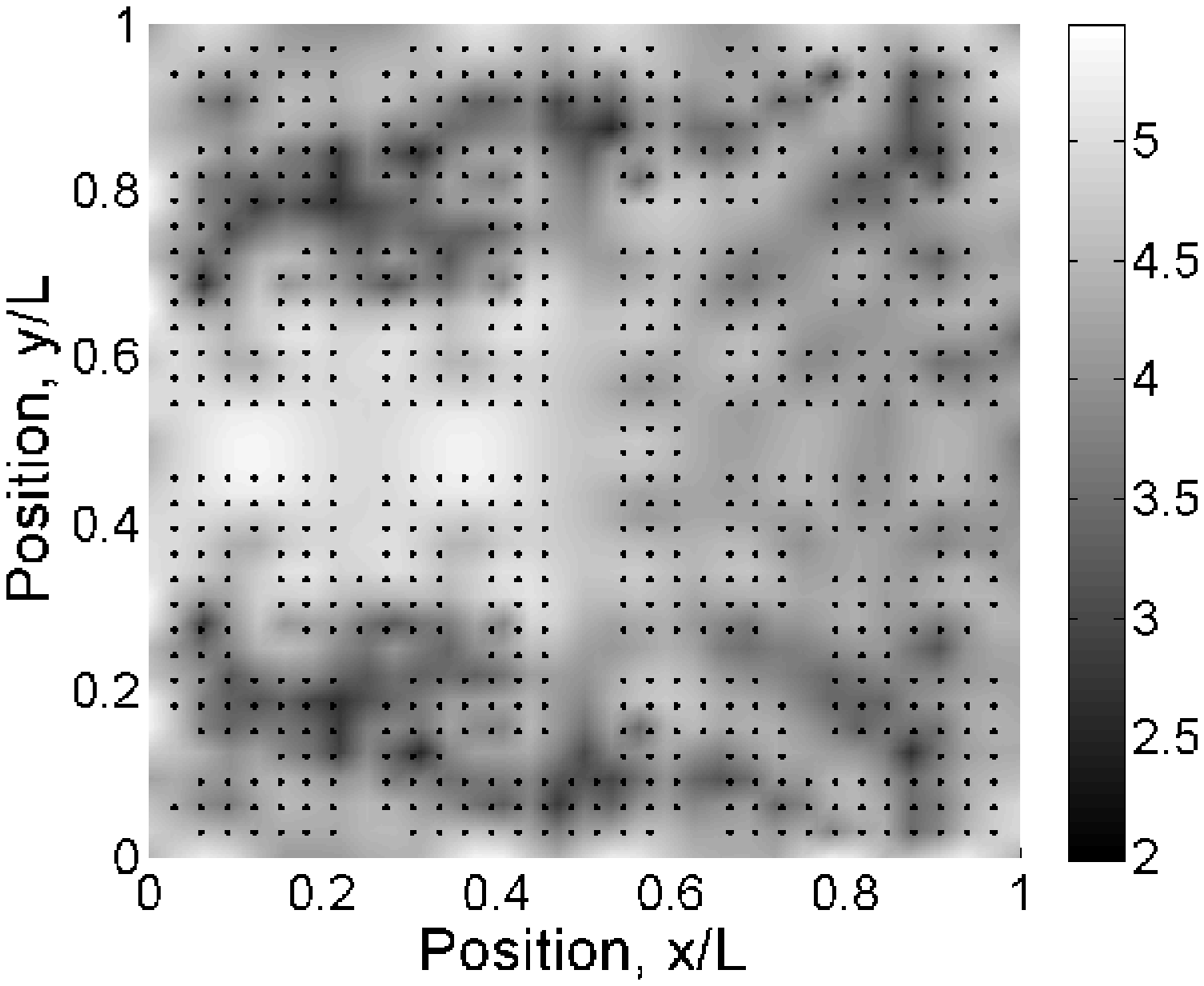}
\caption{{\protect\footnotesize {The 
local field intensity $\log_{10}(|E_{\text {tot}}({\bf r})|^2)$ at the resonant frequency $\omega_1$, corresponding to the induced charge distribution shown in Fig. 4.
 The direction of the  external
field $\mathbf{E}_{\text{ext}}(\protect\omega)$ points along the $y$
axis. The size of the system $L=60$ nm. The dots mark the shape of the trapping potential.}}}
\label{fig6}
\end{figure}
\begin{figure}
\includegraphics[width=5.5cm,angle=0]{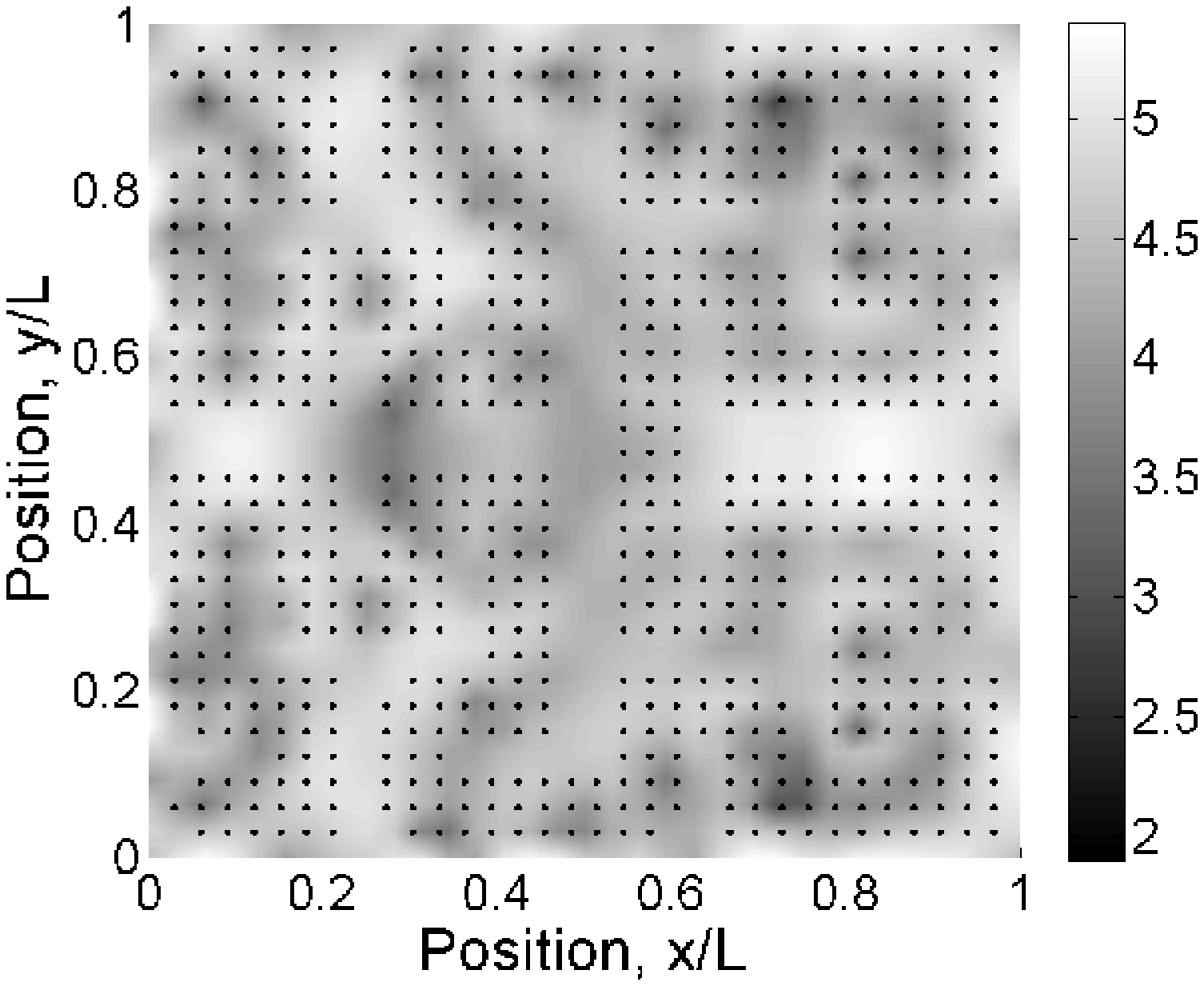}
\caption{{\protect\footnotesize {The 
local field intensity $\log_{10}(|E_{\text {tot}}({\bf r})|^2)$ at the resonant frequency $\omega_2$, corresponding to the induced charge distribution shown in Fig. 5.
 The direction of the  external
field $\mathbf{E}_{\text{ext}}(\protect\omega)$ points along the $y$
axis. The size of the system $L=60$ nm.  The dots mark the shape of the trapping potential.}}}
\label{fig7}
\end{figure}

\begin{figure}[tbp] 
\includegraphics[width=5.5cm,angle=0]{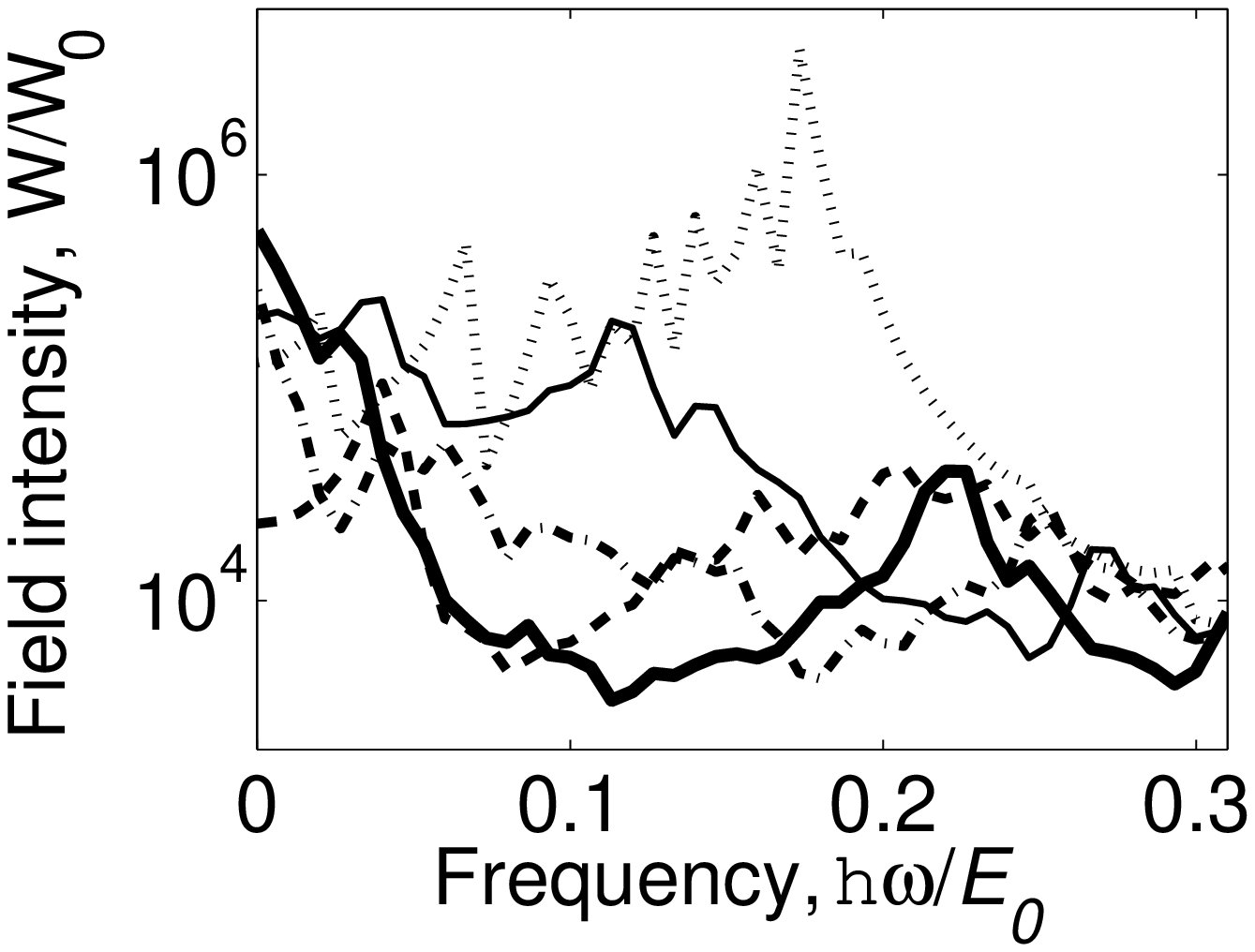}

\caption{{\protect\footnotesize { The normalized local field energy $W/W_0$  as a function of the frequency
 of the external field $\omega$ for different depth of the trapping potential $d$. 
Dotted  line corresponds to  $d=0$, solid thin line:  $d=1 E_0$,  dashed-dotted line  $d=2 E_0$,  dashed line  $d=4 E_0$, solid thick line $d=5 E_0$.
 }}}
\label{fig8}
\end{figure}

First we calculated the ground state electron density for the trapping potential having a constant depth $d$ and a geometry of the Hilbert curve.
We assumed $50$
electrons in the system. 
The potential depth is set to $d=5 E_0$ and has geometry the Hilbert curve  (Fig.1),
The size of the system $60$nm$\times60$nm$\times7.2$nm.   
We have also considered a larger system with the same electron density. 
In Fig.2 we plot the ground state electron density 
assuming $N=200$ electrons in $240$nm$\times240$nm$\times7.2$nm. system (please see Fig 2). 
The potential depth is again set to $d=5 E_0$.  

We calculated the dielectric response of the fractal system shown in Fig. 1 for two electron densities, 
assuming $N=25$ and $N=50$ electrons in the system. In both cases the response has a complex
dependence on the frequency of the external field that is consistent with the complex geometry of 
the nanostructure. 
In the case of $N=25$ the normalized energy of the local field $W/W_0$ in the system 
has a distinct maximum at almost zero frequency (Fig.3, solid line). 
In the case of $N=50$ the response of  the system 
has two clear maxima as a function 
of the frequency of the applied field, one peak is at $\hbar\omega_1\approx0.04 E_0$, and the second is at $\hbar\omega_2\approx0.18 E_0$ 
(Fig.3, dashed line). We have also calculated the dielectric 
response for higher electron densities: $N=100,250$ (not shown). 
Higher carrier concentration leads
to a shorter electron screening length and reduced 
quantum delocalization  effects \cite{ilya1}.   As a result, the dielectric response in systems 
with higher electron densities  is more close to macroscopic (bulk) systems. 
The local field enhancement is localized near  ''corners'' of the Hilbert curve.
The response has many resonances, which are extended over a large 
frequency scale, which is consistent with  the complex self-similar geometry of the nanostructure.
We have investigated the induced charge spatial distribution at resonance frequencies for $N=50$. At $\omega_1$ the induced 
charge density has a  two dimensional profile: the nanostructure is 
divided into upper and lower halves having the induced charge densities of the opposite signs (Fig.4).
At the frequency of the applied field  $\omega_2$ the induced charge spatial distribution has a pronounced one dimensional distribution along the 
path of the Hilbert curve (Fig.5). 
The local field intensities produced by these two charge distributions  are plotted in Figs.(6,7). 
One can see that in the case of quasi one dimensional excitation (Fig.7) the spatial distribution of ''hot spots`` is much more uniform and has 
less inactive area, that beneficial for more reliable SERS measurements.  Note 
that in Figs.(6,7) we used a logarithmic scale.

We have also calculated the dielectric response of the larger nanostructure shown in Fig.2. 
for different depth of the trapping potential $d=0,1,2,4,5 E_0$. Based on the results of simulations shown in Fig. 3, the maximum of the 
response of a larger fractal-like nanostructure should be near $\omega=0$.
In the case of $d=0$ (finite thin film) the system has its response maximum near the surface
 plasmon resonance ($\omega_{pl}=\omega_{bulk}/\sqrt{2}$), where bulk plasmon frequency 
$\omega^2_{bulk}=\frac{\rho e^2}{m^*_e \epsilon_0 \epsilon}$, here $\rho$ is the carrier concentration, 
$e$ is the electron charge, 
$\epsilon_0$ is the vacuum permittivity, and  
$\epsilon\approx 11.1$ is the high-frequency dielectric constant of GaAs \cite{levi}.
With the increasing of the depth $d$ of the trapping potential  
one naively expects that the resonant frequency will shift to a higher frequency, since the same amount of electrons 
will effectively occupy a  smaller volume, that will correspond to
a higher electron density and higher resonant frequency. However, the resonance frequency of the Hilbert geometry has
 moved to  lower values (please, see Fig.8), and at $d=5E_0$ the resonance becomes very close to $\omega=0$.
Note that the overall magnitude of the normalized local field  
is decreased with the increase of the depth $d$. 
One may attribute this to a reduction of the screening in quasi-one dimensional systems. 
As a result, the amplitude of the collective (plasmon) 
resonance in the quasi one dimensional system should be less than in the quasi-two dimensional system.

In summary, we have  studied the dielectric properties of nanoscale systems with the geometry of the Hilbert space filling curve.
In particular, we found that while electrons are trapped in a quasi-one dimensional potential with the Hilbert curve geometry, 
 one can observe either two dimensional or one 
dimensional excitations in the system,  depending on the excitation frequency. 
We have also studied how the plasmon response depends on the depth of the trapping potential, 
which effectively controls the smooth transition of the system geometry from quasi-two dimensional 
to quasi-one dimensional.
We have found that while the average electron density increases with the increasing of the depth of the 
trapping potential, the geometry of the Hilbert curve results in the overall red shift in the plasmon frequency. 
This opens a broad possibility in 
control of the resonance properties of such systems, making them a robust and reproducible 
Raman active nano-structured surfaces for the SERS, and other 
nanoplasmonic applications.


\begin{thebibliography}{99}
\bibitem{shalaev} V. M. Shalaev, E. Y. Poliakov, and V. A. Markel, Phys. Rev. B {\bf 53},
2437 (1996).
\bibitem{sers} S. Nie and S. R. Emory, Science \textbf{275}, 1102 (1997).
\bibitem{stockman} M.I. Stockman, V.M. Shalaev, M. Moskovits, R. Botet, T.F. George, Phys. Rev. B {\bf 46}, 2821 (1992).
\bibitem{stockman1} M. I. Stockman, D. J. Bergman, and T. Kobayashi,
Phys. Rev. B {\bf 69}, 054202 (2004).
\bibitem {raman_frac1}  W. Kim, V. P. Safonov, V. M. Shalaev, and R. L. Armstrong, Phys. Rev. Lett. {\bf 82}, 4811 (1999).
\bibitem {raman_frac2}  M. Montagna, O. Pilla, G. Viliani, V. Mazzacurati, G. Ruocco, and G. Signorelli, Phys. Rev. Lett. {\bf 65}, 1136 (1990).
\bibitem {raman_frac3} T. Keyes and T. Ohtsuki, Phys. Rev. Lett. {\bf 59}, 603 (1987).
\bibitem{nurmikko} T. Atay, J.H. Song, and A.V. Nurmikko, Nano Lett. \textbf{4},1627 (2004).
\bibitem{ilya2} I. Grigorenko, S. Haas, A.
Balatsky and A.  F. J. Levi, New J. Phys. {\bf 10}, 043017 (2008).
\bibitem {lithography_plasmonic_sensors} M. E. Stewart, C. R. Anderton, L. B. Thompson, J. Maria, S. K. Gray,
J. A. Rogers, and R. G. Nuzzo Chem. Rev., {\bf 108}, 494 (2008).
\bibitem{curves} H. Sagan, {\it Space-filling curves}  Springer-Verlag, NY (1994).
\bibitem{ilya1}I. Grigorenko, S. Haas and A.F.J. Levi, Phys. Rev. Lett. {\bf 97}, 036806 (2006);
\bibitem{serp} G. Volpe, G. Volpe, and R. Quidant, Opt. Express., {\bf 19}  3612 (2011).
\bibitem{spie} A.A. Grunin, A.G. Zhdanov and A.A. Fedyanin  Proc. of SPIE {\bf 6728}, 672837, (2007). 
\bibitem{antenna} J. Alda, J. M Rico-Garcia, J. ́M Lopez-Alonso and G Boreman,
 Nanotechnology {\bf 16}, 230 (2005). 

\bibitem{bimetal} C. H. Liu, M. H. Hong, H. W. Cheung, F. Zhang, Z. Q. Huang, L. S. Tan, and
T. S. A. Hor,  Opt. Express. {\bf 16} 10701 (2008). 
\bibitem{manoharan} C. R. Moon, L. S. Mattos, B. K. Foster, G. Zeltzer, W. Ko and H. C. Manoharan,  Science
{\bf  319}, 782 (2008).
\bibitem{levi} J. R. Hayes and A. F. J. Levi, IEEE Journ. of Quant. Elect. {\bf 22}, 1744 (1986).  
\end{thebibliography}
\end{document}